\documentclass[reprint, preprintnumbers, amsfonts, aps, prc, nofootinbib, superscriptaddress]{revtex4-2}

\usepackage{amsmath, amsfonts, amssymb, physics, mathtools}
\usepackage{graphicx}
\usepackage{hyperref}
\usepackage{orcidlink}
\usepackage{xcolor}
\usepackage{float}

\def\l{\left}
\def\r{\right}
\def\mR{\mathcal{R}}

\DeclareMathOperator{\diag}{diag}
\DeclareMathOperator{\Li}{Li}

\begin{document}

\title{Deconfinement and the Charge Susceptibility Matrix of QCD}

\author{Christopher Grace \orcidlink{https://orcid.org/0009-0000-4645-169X}}
\altaffiliation{Both lead authors contributed equally to this work.}

\author{Brady Mowbray
\orcidlink{https://orcid.org/0009-0005-1294-4257}}
\altaffiliation{Both lead authors contributed equally to this work.}

\author{Christopher Plumberg \orcidlink{https://orcid.org/0000-0001-6678-3966}}
\email[Corresponding author: ]{christopher.plumberg@pepperdine.edu}

\affiliation{Natural Science Division, Pepperdine University, Malibu, CA 90263, USA}

%\def\thefootnote{$^\dagger$}
%\footnotetext{All authors contributed equally to this work.}
%\def\thefootnote{\arabic{footnote}}

\begin{abstract}
At sufficiently high temperatures, lattice Quantum Chromodynamics (QCD) calculations predict that nuclear matter undergoes a rapid crossover to a novel, deconfined state known as the quark-gluon plasma (QGP), in which quark flavors essentially behave like independent particles.
Here we explore the nature of this transition by introducing two new thermodynamic observables which we refer to as the \textit{diagonality} and the \textit{rescaled curvature} of the QCD equation of state (EoS).
By comparing these observables in the case of lattice QCD calculations with those of simple toy models and more sophisticated hadronic resonance models, we show that both the diagonality and the rescaled curvature provide complementary insights into the degree of independence between different quark flavors and the underlying degrees of freedom in the system as functions of temperature.
We propose that these observables may help to clarify the nature of the transition region and the onset of deconfinement in nuclear matter.
\end{abstract}

\date{\today}

\maketitle

%===========================================================================
%===========================================================================
%===========================================================================
\section{Introduction}
\label{Sec:Introduction}

Obtaining a thorough understanding of the quantum chromodynamic (QCD) phase diagram remains one of the foremost open challenges in nuclear physics.
At low temperatures and small-to-vanishing baryon densities, QCD matter exists in a confined phase which is well described by a hadron resonance gas (HRG) \cite{Hagedorn:1965st}.
As the temperature increases, QCD matter undergoes a rapid crossover at roughly $T_\mathrm{pc} \approx 155$ MeV to a state in which chiral symmetry is restored and color degrees of freedom are no longer confined within hadrons \cite{Aoki:2006br, Bazavov:2011nk, Borsanyi:2020fev, Borsanyi:2024xrx}.
Increasing temperatures beyond this point eventually results in the formation of an asymptotically free state of quarks and gluons which can be reasonably well described by approaches such as hard thermal loop perturbation theory \cite{Haque:2014rua, Haque:2024gva} or using simplified toy models such as the Stefan-Boltzmann limit \cite{Fodor:2007sy}.
Together with the insights obtained from first principles calculations such as lattice QCD \cite{Ding:2015ona, Nagata:2021ugx, Borsanyi:2021sxv}, these effective models display the rich phase structure of QCD and help to provide a robust description of the behavior of nuclear matter in systems ranging from neutron-star interiors and mergers to relativistic nuclear collisions \cite{Dexheimer:2020zzs, Lovato:2022vgq, Most:2022wgo, Jahan:2026hvs}.

Despite considerable progress in constraining the EoS and phase diagram of nuclear matter, significant questions remain \cite{Du:2024wjm}.
For instance, the intriguing possibility of a third, intermediate `stringy fluid' phase has recently been suggested to exist between the fully confined hadronic phase and the fully deconfined plasma phase for temperatures approximately $T_\mathrm{pc} \lesssim T \lesssim 3T_\mathrm{pc}$ \cite{Fujimoto:2025sxx, Marczenko:2025nhj, Marczenko:2026drt, McLerran:2026dio, Fujimoto:2026syp}.
Such a picture implies a gradual onset of deconfinement which has been shown to be consistent with existing thermodynamic observables such as the Polyakov loop, the trace anomaly, and other bulk quantities \cite{Glozman:2019fku, Glozman:2022lda, Glozman:2022zpy, Cohen:2023hbq, Cohen:2024ffx, Glozman:2025rhe, Glozman:2026glk}, although identifying unique signatures of this novel phase remains an outstanding challenge.
Other perennial challenges include the mapping of the phase diagram at finite conserved charge densities \cite{Nagata:2021ugx, MUSES:2023hyz}, the search for the conjectured QCD critical point \cite{Bluhm:2020mpc}, and the exploration of the low temperature, high baryon density regime \cite{Sorensen:2023zkk, McLerran:2026dio}.

Questions like these underscore the need for new observables which can help to constrain the thermodynamic properties of the QCD EoS across a wide range of temperatures and chemical potentials.
Toward this end, in this paper we introduce two novel observables, both of which are constructed from the full susceptibility matrix $\chi$ of the EoS.
The first observable, called the \textit{diagonality} of $\chi$, measures how close the quark flavor sector of $\chi$ is to a diagonal form and thus quantifies the degree of independence of different quark flavors in the system.

The second observable, called the \textit{rescaled curvature}, relies on the formalism of \textit{thermodynamic geometry}, an approach which employs the machinery of general relativity to characterize the thermodynamic phase space of a system in geometric language \cite{Rao1992, Weinhold:1975xej, Weinhold:1975fyh, Crooks:2007jif}.
Using this formalism, we define the rescaled curvature, showing how it is connected to $\chi$, and demonstrate its sensitivity to the number of conserved charges or quark flavors considered in the system, as well as their degree of independence.
Both observables complement or generalize other thermodynamic quantities which yield insights into flavor and charge correlations and the processes of deconfinement and chiral symmetry restoration in nuclear matter \cite{Koch:2008ia, Ratti:2011au, Bellwied:2019pxh}.
We therefore propose these observables as useful diagnostic tools for understanding the properties of the EoS in different regions of the phase diagram.

This paper is organized as follows.
In Sec.~\ref{Sec:QCDEoS}, we review the standard treatment of the QCD EoS in the grand canonical ensemble and explicitly define the diagonality.
In Sec.~\ref{Sec:ThermodynamicGeometry}, we review the formalism of thermodynamic geometry, an approach which employs the machinery of general relativity to characterize the thermodynamic phase space of a system in geometric language.
Using this formalism, we introduce our second new observable, the rescaled curvature, and show that it connects closely with the number of conserved charges or quark flavors considered in the system, as well as their degree of independence.
In Sec.~\ref{Sec:ToyModelExamples}, we clarify the interpretation of both the diagonality and the rescaled curvature by applying them to simple toy models for which (semi-)analytical results can be easily obtained.
We argue that both observables convey complementary insights into the degrees of freedom of the system and the degree of independence of their respective dynamics.
We then proceed in Sec.~\ref{Sec:Results} to apply the same observables to the more realistic EoSs provided by various hadronic models and lattice QCD.
We find, unsurprisingly, that only lattice QCD reflects the gradual formation of a deconfined medium above the transition temperature $T_\mathrm{pc} \approx 155$ MeV, whereas models which lack a deconfinement mechanism (such as hadronic models) never achieve a completely diagonal $\chi$.
Sec.~\ref{Sec:Conclusions} summarizes our findings and indicates new research directions that our observables could open up.

%===========================================================================
%Other references we should mention:
%\begin{itemize}
%  \item QCD critical surface from constant entropy density \cite{Shah:2026mzu}
  %\item Balance functions (need more than just my papers) \cite{Pratt:2012jz, Pratt:2017oyf, Pratt:2018ebf, Pratt:2019pnd, Pratt:2021xvg, Pratt:2022kvz}
%\end{itemize}

%===========================================================================
%===========================================================================
%===========================================================================
\section{The QCD equation of state and the diagonality}
\label{Sec:QCDEoS_and_BSQDiagonality}

In this section we briefly review the formulation of the QCD equation of state (EoS) of nuclear matter in the grand canonical ensemble (GCE).  We then introduce the \textit{diagonality}, a local, bulk thermodynamic observable designed to quantify the extent to which quark flavors behave independently.

%---------------------------
\subsection{The QCD EoS}
\label{Sec:QCDEoS}

In the GCE, the EoS of nuclear matter can be expressed as a relationship between the pressure $p$ of a system and its temperature $T$ and conserved charge chemical potentials $\mu_B$, $\mu_S$, and $\mu_Q$ for baryon number $B$, net strangeness $S$, and electric charge $Q$, respectively.  These chemical potentials can also be related to those for up, down, and strange ($u$, $d$, and $s$) quark flavors by
\begin{align}
  \mu_u &= \frac{1}{3} \mu_B + \frac{2}{3} \mu_Q \nonumber\\
  \mu_d &= \frac{1}{3} \mu_B - \frac{1}{3} \mu_Q \label{basis_conversion} \\
  \mu_s &= \frac{1}{3} \mu_B - \mu_S - \frac{1}{3} \mu_Q \nonumber
\end{align}
Given $p(T,\vec\mu)$, one can obtain various other thermodynamic quantities by taking appropriate derivatives (or combinations thereof).  For example, the entropy and conserved charge densities are defined by
\begin{align}
  s(T,\vec\mu) \equiv \frac{\partial p}{\partial T},\quad \rho_X(T,\vec\mu) \equiv \frac{\partial p}{\partial \mu_X} \label{thermodynamic_density_definitions}
\end{align}
and the energy density is found using the Gibbs relation
\begin{align}
  \varepsilon(T,\vec\mu) = -p + s T + \sum_X \mu_X \rho_X \label{Gibbs_relation}
\end{align}
Likewise, taking second-order derivatives yields the system's charge susceptibilities, which quantify the fluctuations of conserved charges:
\begin{align}
  \chi_{XY} \equiv \frac{\partial^2 p}{\partial \mu_X \partial \mu_Y} \label{charge_susceptibilities}
\end{align}
Here, $X$ and $Y$ range over some set of independent conserved quantities (e.g., $BSQ$ or $uds$).  We refer to $BSQ$ as the ``charge basis" and $uds$ as the ``flavor basis."

%One is also free to define other observables in terms of these basic derivatives.
%For example, the widely used trace anomaly (sometimes called the ``interaction measure") measures the EoS's degree of non-conformality:
%
%\begin{align}
%  I(T,\vec\mu) &= \frac{\varepsilon - 3p}{T^4} \label{trace_anomaly}
%\end{align}
%
%Notice that $I(T,\vec\mu=0)$ depends only on quantities defined locally at zero chemical potentials, meaning that on its own it contains no direct sensitivity to the degree of non-conformality at finite $\vec\mu$.
%This stands in contrast to the observables we will introduce below, as we see next with the diagonality.

%-----------------------------------
\subsection{The Diagonality}
\label{Sec:Diagonality}

In either the charge basis or the flavor basis, the susceptibility matrix $\chi$ is a symmetric $3 \times 3$ covariance matrix for thermodynamic fluctuations of conserved charges in nuclear matter at a given set of $T$ and $\vec\mu$.
At sufficiently high $T$, QCD predicts that quarks become deconfined, implying that in the flavor basis the off-diagonal susceptibilities $\chi_{ij}$ are expected to become small compared to their diagonal counterparts. 
One therefore expects $\chi$ to become diagonal in the limit of deconfinement. 
This suggests using the degree to which $\chi$ is diagonal as a measure of how close to deconfinement nuclear matter is at a given $T$ and $\vec{\mu}$. 
We refer to this measure as the \textit{diagonality} of $\chi$.

To quantify the diagonality, we first determine the eigenvalues $(\lambda_1, \lambda_2, \lambda_3)$ of $\chi$ (in the flavor basis) and then define the diagonal matrix $\chi^{(D)} \equiv \diag\l( \lambda_1, \lambda_2, \lambda_3 \r)$.\footnote{We note that numerical diagonalization algorithms do not generally prescribe a unique ordering of the eigenvalues.
In order to ensure an unambiguous value for the diagonality, we choose the ordering of the $\lambda_i$ to be the one which minimizes the Frobenius norm of the resulting deviation matrix $\delta\chi$.}
If $\chi$ were perfectly diagonal, then we would have $\chi = \chi^{(D)}$ identically.
We can therefore quantify $\chi$'s deviation from diagonality by considering the average size of the elements in the deviation matrix $\delta \chi \equiv \chi - \chi^{(D)}$.
To do this, we compute the Frobenius norm of $\delta \chi$, which is defined for any $N \times N$ matrix $M$ by
\[\l|\l| M \r|\r|_F \equiv \sqrt{\sum_{i=1}^N \sum_{j=1}^N \l| M_{ij} \r|^2}\]
We then define the diagonality $\Delta_\nu$ of $\chi$ to be the dimensionless ratio
\begin{align}
  \Delta_\nu
    \equiv \l(\frac{\l|\l| \chi - \chi^{(D)} \r|\r|_F}{\l|\l| \chi^{(D)} \r|\r|_F}\r)^\nu
    \equiv \l(\frac{\l|\l| \delta\chi \r|\r|_F}{\l|\l| \chi^{(D)} \r|\r|_F}\r)^\nu. \label{Delta_definition}
\end{align}
where $\nu$ is a positive real number.
Note that the limit $\Delta_\nu \to 1$ corresponds to a matrix with all equal elements, while $\Delta_\nu \to 0$ implies a perfectly diagonal $\chi$, for any $\nu > 0$.

Intuitively, $\Delta_\nu$ represents the importance of off-diagonal susceptibilities compared to diagonal ones.
Since by definition $\chi$ is a covariance matrix, its entries are \textit{quadratic} in locally fluctuating conserved charges and its (off-)diagonal elements measure the (co)variances of these fluctuations.
However, in considering the degree of independence of different quark flavors, it may perhaps be more intuitive to employ a measure which is \textit{linear} in the charge fluctuations, rather than quadratic.
This can be done by fixing $\nu = 1/2$, which we will do for the rest of this paper.
We will henceforth drop the subscript and write simply $\Delta$ where no confusion can arise.
Some \textit{a posteriori} justification for this choice will be seen below, where we will show that $\Delta = 0.5$ (for this choice of $\nu$) nearly coincides with the widely accepted value for the chiral crossover temperature $T_c \approx 156.5$ (1.5) MeV \cite{HotQCD:2018pds} when one evaluates $\Delta$ for lattice QCD \cite{Noronha-Hostler:2019ayj}.

\subsection{Models of the QCD EoS}

In this study we consider three different models of the QCD pressure:
  (i), a parametrization of lattice QCD including the three conserved charges $B$, $S$, and $Q$ \cite{Noronha-Hostler:2019ayj};
  (ii), the Ideal hadronic resonance gas (HRG) model which treats nuclear matter as a quantum statistical gas of non-interacting hadrons \cite{Vovchenko:2016rkn}; and
  (iii), the Quantum van der Waals (QvdW) HRG model which incorporates both attractive and repulsive interactions into the hadronic system \cite{Vovchenko:2015vxa, Vovchenko:2016rkn}.
The latter two models are studied using Thermal-FIST \cite{Vovchenko:2019pjl}.
Below, we briefly summarize the main features of each model which are relevant to our analysis.
For completeness, we also include a brief discussion of the Stefan-Boltzmann limit of nuclear matter representing a gas of massless, non-interacting quarks and gluons, as well as a `minimal' set of hadrons or \textit{minimal hadron list} (MHL) which we will employ, in the Boltzmann approximation, in order to clarify the effect of varying particle content on the observables below.
%These will help to provide intuition for the observables we introduce below.

\subsubsection{Lattice QCD}

Since QGP formation is expected only in the lattice QCD parameterization, we expect that $\chi$ should only approach a diagonal form in that model at sufficiently large $T$.
We do not expect the HRG models to give a completely diagonal $\chi$ in the flavor basis, since quarks remain confined within hadrons for all $T$ in these models and hence cross-flavor correlations should always be non-vanishing.

Here, we employ the parameterization constructed in \cite{Noronha-Hostler:2019ayj} to extend the lattice QCD results \cite{Borsanyi:2018grb} to finite $B$, $S$, and $Q$.
The form of the pressure is defined by a Taylor expansion in $\mu_{B,S,Q}/T$:
\begin{align}
  \frac{p(T,\mu_B,\mu_S,\mu_Q)}{T^4} = \sum_{i,j,k} \frac{\chi^{BSQ}_{ijk}(T)}{i! j! k!} \l( \frac{\mu_B}{T} \r)^i \l( \frac{\mu_S}{T} \r)^j \l( \frac{\mu_Q}{T} \r)^k \label{lattice_pressure}
\end{align}
where $i+j+k \leq 4$ and each coefficient $\chi^{BSQ}_{ijk}(T)$ smoothly matches a Pade approximant of lattice results \cite{Borsanyi:2018grb} in the temperature range $135 \leq T \leq 220$ MeV to an HRG model at $T \lesssim 135$ MeV \cite{Alba:2017mqu}.
Above $T = 220$ MeV, the same coefficients are extrapolated toward their respective Stefan-Boltzmann limits.
This EoS and others like it \cite{Monnai:2019hkn, Monnai:2021kgu, Monnai:2024pvy} have been used in hydrodynamic simulations of nuclear collisions which incorporate dynamical evolution for multiple conserved charges \cite{Karpenko:2013wva, Plumberg:2024leb, JETSCAPE:2025wac, Pala:2025qoa}.

\subsubsection{Ideal HRG}

For the Ideal HRG, all hadronic species are non-interacting and so the total pressure is simply a sum over pressures of each individual species:
\begin{align}
  \frac{p^\mathrm{Id}}{T^4}
    &\equiv \sum_{i\in\text{hadrons}} p_i \\
    &= \sum_{i\in\text{hadrons}} \frac{g_i m_i^2 T^2}{2\pi^2} \sum_{n = 1}^\infty \frac{(\pm 1)^{n+1} \lambda_i^n}{n^2} K_2\l( \frac{n m_i}{T} \r) \label{p_Ideal_HRG}
\end{align}
where the `+' (`-') sign is for fermions (bosons), and $p_i$, $m_i$, and $g_i$ are the partial pressure, mass, and spin degeneracy of the $i$th hadron, respectively.
Each hadron is also indexed by its baryon number $B_i$, strangeness $S_i$, and electric charge $Q_i$, in terms of which its fugacity $\lambda_i$ is defined by
\begin{align}
  \lambda_i &\equiv e^{\mu_i / T}
\end{align}
with
\begin{align}
  \mu_i &\equiv \mu_B B_i + \mu_S S_i + \mu_Q Q_i
\end{align}
the chemical potential of the $i$th hadron species.
If the sum over $n$ is truncated after the first term, one further obtains the well-known Boltzmann approximation, which effectively suppresses quantum effects and thereby treats both bosons and fermions as classical particles.
We will use this approximation below in our discussion of the MHL as a toy model.

It will be convenient below to evaluate the Ideal susceptibility $\chi_i$ for a single hadronic species $i$ with spin $s$ in the high-temperature limit.
We will be interested specifically in the susceptibility when $\mu_i=0$.
This can be obtained by differentiating $p_i$ in \eqref{p_Ideal_HRG} twice with respect to $\mu_i$, setting $\mu_i = 0$, taking $T \to \infty$ and finally evaluating the infinite sum over $n$.
One finds that
\begin{align}
  \chi_i(T,\mu_i = 0) &= \l.\frac{\partial^2 p_i}{\partial \mu_i^2}\r|_{\mu_i = 0} \\
  &= \frac{2}{\pi^2}(2s+1)(-1)^{2s} T^2 \Li_2\l( (-1)^{2s} \r)
\end{align}
which, for $s=0$ or $s=1/2$, reduces to $T^2/3$, as one may easily check by evaluating \eqref{p_Ideal_HRG} numerically \cite{Koch:2008zz}.
Hence the hadron susceptibilities at $\mu_i = 0$ are equal for $s=0$ and $s=1/2$.
We will make use of this fact below.

\subsubsection{QvdW HRG}

The QvdW HRG \cite{Vovchenko:2015vxa, Vovchenko:2016rkn, Vovchenko:2017zpj} can be formulated as an extension of the Ideal HRG in the spirit of the traditional van der Waals equation of state, whose pressure is usually written in the canonical ensemble in terms of the temperature and the density as
\begin{align}
  p(T, \l\{n_k\r\}) &= \sum_i p^{\mathrm{Id}} \l( T, \frac{n_i}{f_i} \r) - \sum_{i,j} a_{ij} n_i n_j \label{vdWEoS}
\end{align}
where $f_i = 1 - \sum_j \tilde{b}_{ji} n_j$.
Translating to the grand canonical ensemble leads to the following system of transcendental equations which must be solved numerically:
\begin{align}
  p_i^* &\equiv p^\mathrm{Id}(T,\mu_i^*), \quad n_i^* \equiv n^\mathrm{Id}(T,\mu_i^*) \\
  p(T, \l\{\mu_k\r\}) &= \sum_i p_i^* - \sum_{i,j} a_{ij} n_i n_j \\
  n_i^* &= \frac{n_i(T, \l\{\mu_k\r\})}{f_i} \\
  \mu_i^* &= \mu_i - \sum_j \tilde{b}_{ij} p_j^* + \sum_j (a_{ij} + a_{ji}) n_j(T, \l\{\mu_k\r\}),
\end{align}
and
\begin{align}
  n_i(T, \l\{\mu_k\r\}) &= \frac{\partial p}{\partial \mu_i}.
\end{align}
The coefficients $a_{ij}$ and $\tilde{b}_{ij}$ allow for attractive and repulsive interactions, respectively, between different hadron species.  In this paper, we assume that these coefficients are nonzero only for baryon-baryon (or antibaryon-antibaryon) pairs, in which case they are given by the values
\begin{align}
  a_{ij} &= 329 \text{ MeV fm}^3,\quad \tilde{b}_{ij} = b_0 \equiv 3.42 \text{ fm}^3
\end{align}
To assess the role of the excluded volume parameter $b_0$ in our results below, we will also in some cases plot results with $\tilde{b}_{ij} = 0.5b_0$ and $\tilde{b}_{ij} = 2b_0$ for comparison.

\subsubsection{Stefan - Boltzmann limit}

For completeness, we also record the Stefan-Boltzmann limit for an arbitrary number of non-interacting quarks and gluons in the high-$T$ regime.  For $N_c$ colors and $N_f$ quark flavors, one finds \cite{Ratti_2021}:
\begin{align}
  p &= \frac{\pi^2}{45} T^4\l( N_c^2 - 1 \r) \theta \nonumber\\
    &+ \frac{N_c}{3} \sum_{f=1}^{N_f}
             \l( \frac{7\pi^2 T^4}{60}
                + \frac{\mu_f^2 T^2}{2}
                + \frac{\mu_f^4}{4\pi^2} \r) \\
    &= T^4 \l[
       \frac{8\pi^2}{45} \theta
          + \sum_{f=1}^{N_f}
             \l( \frac{7\pi^2}{60}
                + \frac{1}{2} \l( \frac{\mu_f}{T} \r)^2
                + \frac{1}{4\pi^2} \l( \frac{\mu_f}{T} \r)^4
             \r)
           \r] \label{SB_limit_massless}
\end{align}
where the second step sets $N_c = 3$.  The results presented in this paper take $N_f = 3$ in the $uds$ flavor basis, making \eqref{SB_limit_massless} a function of $T$, $\mu_u$, $\mu_d$, and $\mu_s$.  The dependence on $\mu_{B/S/Q}$ can be restored by making use of Eq.~\eqref{basis_conversion}.
Here for convenience we have also introduced a quantity $\theta = 0$ or 1 which controls whether or not gluons are included in the EoS.
We will explain the reason for including this below.

We note additionally that the Stefan-Boltzmann limit is a special instance of the more general \textit{conformal limit}, for which one has
\begin{align}
  \varepsilon &= 3 p
\end{align}
This is automatically satisfied by any EoS of the form
\begin{align}
  p = T^4 f\l( \frac{\mu_1}{T}, \frac{\mu_2}{T}, \frac{\mu_3}{T}, \ldots \r) \label{EoS_conformal_limit_general}
\end{align}
Such forms generally arise in scale invariant theories and need not imply weak coupling \cite{Panero:2009tv}.

\subsubsection{Minimal hadron list}
\label{Sec:MHL_EoS}

HRG models conventionally employ extensive lists of up to several hundred known hadron resonances in order to attain a semi-quantitative description of nuclear matter at temperatures below the chiral restoration temperature.
In general, those lists which include a more comprehensive set of such resonances thereby provide a more accurate description of nuclear matter in this regime.
Here, we introduce what we call a ``minimal hadron list" (MHL) of hadrons, which we define to be a minimum collection of hadrons which carry linearly independent combinations of the conserved charges ($B$, $S$, and $Q$) under investigation.
Such a list is not unique and need not provide a quantitative description of nuclear matter.
Here, we will use an MHL consisting of $\pi^\pm$, $K^\pm$, and $p/\bar{p}$ as a highly simplified toy model of nuclear matter.

We will apply the Ideal and QvdW HRG models to the MHL particle list below in Sec.~\ref{Sec:Results}.  In this section we discuss an additional simplication in the form of the Boltzmann approximation, which will permit us to keep the analysis in Sec.~\ref{Sec:ToyModelExamples} analytically tractable.
We stress that this is not intended as an accurate description of nuclear matter, but simply meant to build intuition for the kind of information which the diagonality and thermodynamic curvature contain about the EoS.

With these simplifications, together with the fact that $g_\pi = g_K = 1$ and $g_p = 2$, the Ideal pressure of the MHL reduces to
\begin{align}
  p &= p_\pi + p_K + p_p \\
    &= \frac{T^2}{\pi^2}
       \l[ m_\pi^2 \cosh\l( \frac{\mu_Q}{T} \r) K_2\l( \frac{m_\pi}{T} \r) \r. \nonumber\\
    & \l. \qquad \qquad + m_K^2 \cosh\l( \frac{\mu_S + \mu_Q}{T} \r) K_2\l( \frac{m_K}{T} \r) \r. \nonumber\\
    & \l. \qquad \qquad + 2 m_p^2 \cosh\l( \frac{\mu_B + \mu_Q}{T} \r) K_2\l( \frac{m_p}{T} \r) \r] \label{pi_K_p_EoS}
\end{align}
The charge susceptibility matrix is
\begin{align}
  \chi &= \frac{1}{T^2}
    \begin{pmatrix}
      p_p & 0   & p_p \\
      0   & p_K & p_K \\
      p_p & p_K & p_\pi + p_K + p_p
    \end{pmatrix} , \label{pi_K_p_chi_general}
\end{align}
with partial pressures ($p_\pi$, $p_K$, and $p_p$) defined implicitly in \eqref{pi_K_p_EoS}.  In the limit as $T \to \infty$ with $\mu_i/T$ held fixed, \eqref{pi_K_p_EoS} simplifies still further to
\begin{align}
  p &= \frac{2T^4}{\pi^2} \l[\cosh\l( \frac{\mu_Q}{T} \r)
           + \cosh\l( \frac{\mu_S + \mu_Q}{T} \r) \r. \nonumber\\
    & \l. \qquad \qquad + 2 \cosh\l( \frac{\mu_B + \mu_Q}{T} \r)
       \r] \label{pi_K_p_EoS_massless}
\end{align}
As with the Stefan-Boltzmann limit \eqref{SB_limit_massless}, Eq.~\eqref{pi_K_p_EoS_massless} is again easily seen to be an instance of the conformal limit.

%===========================================================================
%===========================================================================
%===========================================================================
\section{Thermodynamic geometry}
\label{Sec:ThermodynamicGeometry}

Thermodynamic geometry (TG) is an approach to characterizing a system's thermodynamic properties by associating a metric to the space of thermodynamic states available to the system in question \cite{Rao1992, Weinhold:1975xej, Weinhold:1975fyh, Crooks:2007jif}.
The approach itself has been in existence for a number of decades and has been applied to a wide variety of different systems, including those with and without critical behavior.
There has also been some recent attention paid to exploring the consequences of thermodynamic geometry in nuclear matter \cite{Castorina:2018gsx, Castorina:2018ayy, Castorina:2019jzw, Zhang:2019neb, Murgana:2023pyx, Oshima:2026bub}, although it has not yet been widely adopted by the community.
Here, we generalize the TG approach to the full $BSQ$ susceptibility matrix which is the focus of the present work.

In this section, we review only the essential formalism of thermodynamic geometry needed for our study below.
We also briefly outline some of the main ways of interpreting this formalism.
More complete reviews are available in Refs.~\cite{Ruppeiner:1979bcp, Ruppeiner:1995zz, Ruppeiner:2010dzw, Ruppeiner:2011gm}.

%===========================
\subsection{Formalism}
\label{Sec:Formalism}

One way of making the connection between thermodynamics and geometry begins by introducing the following thermodynamic coordinates (in the so-called ``entropy representation"):
\begin{align}
  (x^0, x^1, x^2, x^3) \equiv \l( \beta, -\vec{\gamma} \r) \equiv \l( \frac{1}{T}, -\frac{\vec{\mu}}{T} \r) \label{TG_coordinates}
\end{align}
where $\vec{\mu} = (\mu_B, \mu_S, \mu_Q)$ and $\vec{\gamma}$ represents the corresponding logarithmic fugacities.
  Introducing the thermodynamic potential\footnote{Although other thermodynamic potentials are available \cite{Ruppeiner:1995zz}, we adopt Eq.~\eqref{thermodynamic_potential_definition} here since it has historically been the usual choice for analyses applying thermodynamic geometry to the EoS of nuclear matter \cite{Castorina:2018ayy, Castorina:2018gsx, Castorina:2019jzw, Castorina:2020vbh}.  For an alternative choice which is often more convenient when using the grand canonical ensemble, see Appendix \ref{App:RicciCurvatureGeneral}.}
\begin{align}
  \phi \equiv \frac{P}{T} \label{thermodynamic_potential_definition}
\end{align}
one then defines a metric on the thermodynamic phase space by $g_{\alpha\beta} \equiv \partial^2\phi / \partial x^\alpha \partial x^\beta \equiv \phi_{,\alpha\beta}$, where the last equivalence uses the standard ``comma notation" as shorthand for partial derivatives.
For instance, in this notation we have
\begin{align}
  g_{02} = \phi_{,02} = \frac{\partial^2 \phi}{\partial \beta \partial (-\gamma_S)}
\end{align}
Note that the metric constructed for the coordinates in \eqref{TG_coordinates} is four dimensional (4D).
However, one is also free to consider a subset of chemical potentials, corresponding to the coordinates of a two-dimensional (2D) metric.
For example, if one uses $\mu_B$ only, the 2D coordinates are
\begin{align}
  (x^0, x^1) \equiv \l( \beta, -\gamma_B \r) \equiv \l( \frac{1}{T}, -\frac{\mu_B}{T} \r) \label{TG_coordinates_2D}
\end{align}
In the 2D case, the metric itself is then
\begin{align}
  g_{2D} \equiv
    \begin{pmatrix}
      \phi_{,00} & \phi_{,01} \\
      \phi_{,01} & \phi_{,11}
    \end{pmatrix} \label{metric2D_definition}
\end{align}
and in the 4D case,
\begin{align}
  g_{4D} \equiv
    \begin{pmatrix}
      %---------------------------------------------------------------
      \phi_{,00} & \phi_{,01} & \phi_{,02} &  \phi_{,03} \\
      %---------------------------------------------------------------
      \phi_{,01} & \phi_{,11} & \phi_{,12} &  \phi_{,13} \\
      %---------------------------------------------------------------
      \phi_{,02} & \phi_{,12} & \phi_{,22} &  \phi_{,23} \\
      %---------------------------------------------------------------
      \phi_{,03} & \phi_{,13} & \phi_{,23} &  \phi_{,33}
      %---------------------------------------------------------------
    \end{pmatrix} \label{metric4D_definition}
\end{align}
With the metric in hand, one can then employ the usual machinery of Riemannian geometry to analyze the corresponding structure of the thermodynamic phase space.
We adopt the following standard definitions \cite{weinberg1972gravitation}:
\begin{align}
  & \text{Christoffel symbols:} \nonumber\\
  & \qquad \Gamma^\lambda_{\mu\nu} \equiv \frac{1}{2} g^{\lambda \sigma} \l( g_{\mu\sigma,\nu} + g_{\nu\sigma,\mu} - g_{\mu\nu,\sigma} \r) \label{Christoffel_definition} \\
  & \text{Riemann tensor:} \nonumber\\
  & \qquad R^\lambda_{\mu\nu\rho} \equiv \partial_\nu \Gamma^\lambda_{\mu\rho} - \partial_\rho  \Gamma^\lambda_{\mu\nu} + \Gamma^\sigma_{\mu\rho} \Gamma^\lambda_{\sigma\nu} - \Gamma^\sigma_{\mu\nu} \Gamma^\lambda_{\sigma\rho} \label{Riemann4D_definition} \\
  & \text{Ricci tensor:} \nonumber\\
  & \qquad R_{\mu\nu} \equiv R^\lambda_{\mu\lambda\nu}
    = \partial_\lambda \Gamma^\lambda_{\mu\nu} - \partial_\nu  \Gamma^\lambda_{\mu\lambda} + \Gamma^\sigma_{\mu\nu} \Gamma^\lambda_{\sigma\lambda} - \Gamma^\sigma_{\mu\lambda} \Gamma^\lambda_{\sigma\nu}  \label{Ricci2D_definition} \\
  & \text{Ricci scalar:} \nonumber\\
  & \qquad R \equiv g^{\mu\nu} R_{\mu\nu} \label{RicciScalar_definition}
\end{align}
In the 2D case, the result for the Ricci scalar $R$ simplifies considerably to \cite{Janyszek:1990wdh}
\begin{align}
  R &\equiv \frac{1}{2 |g|^2} \l| \begin{matrix}
                                \phi_{,00} & \phi_{,01}  & \phi_{,11} \\
                                \phi_{,000} & \phi_{,001}  & \phi_{,011} \\
                                \phi_{,001} & \phi_{,011}  & \phi_{,111}
                              \end{matrix} \r|\,, \label{RicciScalar2D_definition}
\end{align}
where $|g| \equiv \det(g_{\mu\nu})$ is the determinant of the metric.
In the 4D case, $R$ becomes vastly more complicated and can only be evaluated analytically in very simple cases (some of which we consider below).
In all other cases we evaluate $R$ numerically.
Note that $R$ has units of volume, which we report here in $\text{fm}^3$.

%===========================
\subsection{Interpretation}
\label{Sec:Interpretation}

While in principle a great deal of information about a metric's curvature is contained within the components of its Riemann tensor, in this study we focus only on the scalar curvature $R$, which distills the thermodynamic structure to a single number.
One proposed way of interpreting $R$ is as a measure of interactions in the system: $|R|$ corresponds to the correlation length, while the sign of $R>0$ (or $R<0$) indicates whether interactions are predominantly repulsive (or attractive, respectively) \cite{Ruppeiner:2010dzw}.
In this context, both models with a critical point and those containing bosonic degrees of freedom have generally been found to correspond to $R<0$ (as these tend to produce clustering effects which mimic attractive interactions), whereas fermionic systems generically have $R>0$.
Divergences in $R$ usually coincide with phase transitions.

The picture that $R$ measures interactions is complicated somewhat by systems, such as the classical multicomponent ideal gas, which lack interparticle interactions entirely but nevertheless possess small and non-vanishing scalar curvatures on the order of the specific volume \cite{PhysRevA.41.2200}.
In fact, only in the limit of a single component does the ideal gas exhibit zero curvature \cite{Nulton_1985}.
This suggests that $R$ measures something other than the mere strength of interactions \cite{Rodrigo:2024rnj}.
The scalar curvature can also be evaluated for black holes when viewed as thermodynamic systems, although the interpretation in this case remains unclear \cite{Ruppeiner:2013yca, Ruppeiner_2013, e16126515}.

While the precise interpretation of $R$ continues to be debated \cite{PhysRevE.106.054141}, it is generally agreed that the scalar curvature conveys non-trivial insights into the phase space structure of thermodynamic systems.
In this paper, one of our goals is to explore how the thermodynamic geometry of nuclear matter in the QCD phase structure reflects the presence of multiple conserved charges and, in particular, in what ways it may provide complementary information to the diagonality introduced above.

In order to build some intuition for the interpretation of $\Delta$ and $R$, in Sec.~\ref{Sec:ToyModelExamples} we consider the behavior of these quantities for two simple equations of state, in order to establish an intuitive baseline for interpreting the more physically realistic thermodynamic geometry predicted by lattice QCD.
The curvature $R$ has already been evaluated for a variety of different models of nuclear matter, including perturbative approaches to hot QCD \cite{Bellucci:2010ss}, the hadron resonance gas \cite{Castorina:2018ayy, Castorina:2018gsx}, the Nambu-Jona Lasinio model \cite{Castorina:2019jzw}, the quark-meson model \cite{Castorina:2020vbh}, and the functional renormalization group \cite{Murgana:2023pyx}.
In addition, lattice QCD itself has also been studied geometrically, although this has so far been done only for a single conserved charge chemical potential at a time (typically $\mu_B$).
One goal of this study is to analyze lattice QCD geometrically in the presence of multiple conserved charges $BSQ$.

%(*\textit{From the Introduction}*)
%The Ricci curvature has already been proposed as a measure of quark deconfinement, although recent analyses \cite{Castorina:2018ayy, Murgana:2023pyx, Oshima:2026bub} have been limited to the ($T$,$\mu_B$)-plane and have not included $\mu_S$ and $\mu_Q$ as independent directions, rendering the QCD phase diagram as a 2D plane rather than a fully 4D space \cite{XXX}.

%===========================
\section{Toy Model Examples}
\label{Sec:ToyModelExamples}

In this section we evaluate the quantities $\Delta$ and $R$ for the two toy models considered above: (i), a non-interacting gas of massless quarks and gluons; and (ii), a minimal system of hadrons ($\pi^\pm$, $K^\pm$, and $p/\bar{p}$) in the Boltzmann approximation.  %For the scalar curvature $R$, we note that one is free to use the complete set of chemical potentials ($\mu_B$, $\mu_S$, $\mu_Q$) or only a subset of them

\subsubsection*{Case 1: Massless quarks and gluons}

We first consider the simplest case of massless, non-interacting quarks and gluons.
We begin by discussing the 2D case with $\mu_S = \mu_Q = 0$ and retaining $(T,\mu_B)$ as variables.%
\footnote{
  We refer to this choice as the ``unconstrained" case.
  A different 2D `projection' of the full 4D QCD EoS can be obtained instead by imposing the constraints $\rho_S = 0$ and $\rho_Q = 0.4 \rho_B$ at any given $T$ and $\mu_B$, as an approximation to the conditions which prevail on average in nuclear collisions \cite{Braun-Munzinger:1995uec, Braun-Munzinger:1999hun, Cleymans:2005xv}, and then solving these constraints numerically for $\mu_S$ and $\mu_Q$ as implicit functions of $T$ and $\mu_B$.
  This ``constrained" case was studied and compared to the ``unconstrained" case in Ref.~\cite{Castorina:2018ayy}.  The differences were found to have only negligible effects on the thermodynamic curvature.
}
Inserting the relations \eqref{basis_conversion} into \eqref{SB_limit_massless} one finds for $\theta=1$, that
\begin{align}
  R
    = \frac{3\l(513 \pi^6 T^6 + 171 \pi^4 T^4 \mu_B^2 + 11 \pi^2 T^2 \mu_B^4 + \mu_B^6\r)}
           {T\l( 57 \pi^4 T^4 + 10\pi^2 T^2 \mu_B^2 + \mu_B^4 \r)^2}
\end{align}
and for $\theta=0$
\begin{align}
  R
    = \frac{15 \l(5 \mu_B^6 + 1701 \pi^6 T^6 + 567 \pi^4 \mu_B^2 T^4 - 9 \pi^2 \mu_B^4 T^2\r)}
           {T \l(5 \mu_B^4 + 189 \pi^4 T^4 + 18 \pi^2 \mu_B^2 T^2 \r)^2}
\end{align}
Using \eqref{thermodynamic_density_definitions} one may also easily compute the entropy and baryon densities, which are (for $\theta=1$)
\begin{align*}
  s(T,\mu_B) &= \frac{1}{9}\l( 19 \pi^2 T^3 + 3T\mu_B^2 \r) \\
  \rho_B(T,\mu_B) &= \frac{\mu_B}{27\pi^2}\l( 9 \pi^2 T^2 + \mu_B^2 \r)
\end{align*}
and (for $\theta=0$)
\begin{align*}
  s(T,\mu_B) &= T\l( \frac{7\pi^2}{5}T^2 + \frac{1}{3} \mu_B^2 \r),
\end{align*}
while $\rho_B$ is the same as for the $\theta = 1$ case.
Noting that the entropy density has units of 1/fm$^3$ while $R$ has units of fm$^3$, it is natural to consider their dimensionless product, which we designate the \textit{rescaled (thermodynamic) curvature} $\mR$ of the EoS:
\begin{align}
  \mR &\equiv s R \label{mR_definition}
\end{align}
Remarkably, the rescaled curvature at $\mu_B = 0$ reduces to a simple constant:
\begin{align}
  \text{2D case: } \mR(T,0) = 1 \label{rescaled_R_mu=0_2D}
  \end{align}
Crucially, this relation holds regardless of the value of $\theta$.  In other words, the product of $s$ and $R$ is the same in the Stefan-Boltzmann EoS \eqref{SB_limit_massless}, whether or not gluons are included in the system.

The 4D case can be handled in a similar way.
In this case, the general expression for $R$ for arbitrary $\vec\mu$ is very complicated and we do not reproduce it here.
Setting $\vec\mu = 0$, however, again yields a constant value:
\begin{align}
  \text{4D case: } \mR\l(T,\vec\mu=0\r) = \frac{15}{2}  \label{rescaled_R_mu=0_4D}
\end{align}
As in the 2D case, this result is independent of the value of the parameter $\theta$.
In general, one can show (see App.~\ref{App:RicciCurvatureGeneral}) that the rescaled curvature at $\vec\mu = 0$ is given for $N_f$ massless, non-interacting quark flavors by
\begin{align}
  &\text{$(N_f+1)$-D case: } \nonumber\\
  & \quad \mR_\mathrm{C} \equiv \mR\l(T,\vec\mu=0\r) = \frac{1}{4} N_f\l( 3 N_f + 1 \r) ,\label{rescaled_R_mu=0_ND_SF}
\end{align}
regardless of whether gluons are included in the EoS.
More generally, the rescaled curvature gives a way to quantify how well an EoS may be described by a gas with some number of non-interacting quark flavors or conserved charges.

Eq.~\eqref{rescaled_R_mu=0_ND_SF} is one of this paper's primary results.
It indicates that the quantity $\mR$ at $\vec\mu = 0$ provides a measure of the degree of independence of quark flavors which is insensitive to the importance of the gluonic contribution.  This should be contrasted with the similar quantity $s/T^3$, which functions as a measure of the effective number of degrees of freedom in the system and is affected by the presence of both quarks and gluons \cite{McLerran:2026dio}.
Note further that $\mR$ is generally only constant at $\vec\mu = 0$; at finite density $s$ and $R$ differ in their functional forms and so their dependences on $T$ and $\vec\mu$ no longer cancel.
It would be interesting to consider generalizations of Eq.~\eqref{mR_definition} to include a dependence also on the charge densities $\rho_X$, but we do not consider this possibility here.
We note furthermore that $R(T,\vec\mu = 0) > 0$ for all $T$ in both the 2D and 4D cases of the non-interacting massless limit.
This behavior should be contrasted with that of the trace anomaly, which vanishes identically for any conformal system, regardless of the number of massless non-interacting degrees of freedom it contains or how close it is to deconfinement.

Finally, since a non-interacting gas of quarks and gluons is by definition diagonal, it follows trivially that the diagonality in this case is exactly zero.

\subsubsection*{Case 2: Minimal Hadron List}

\begin{figure}
  \centering
  \includegraphics[width=\linewidth]{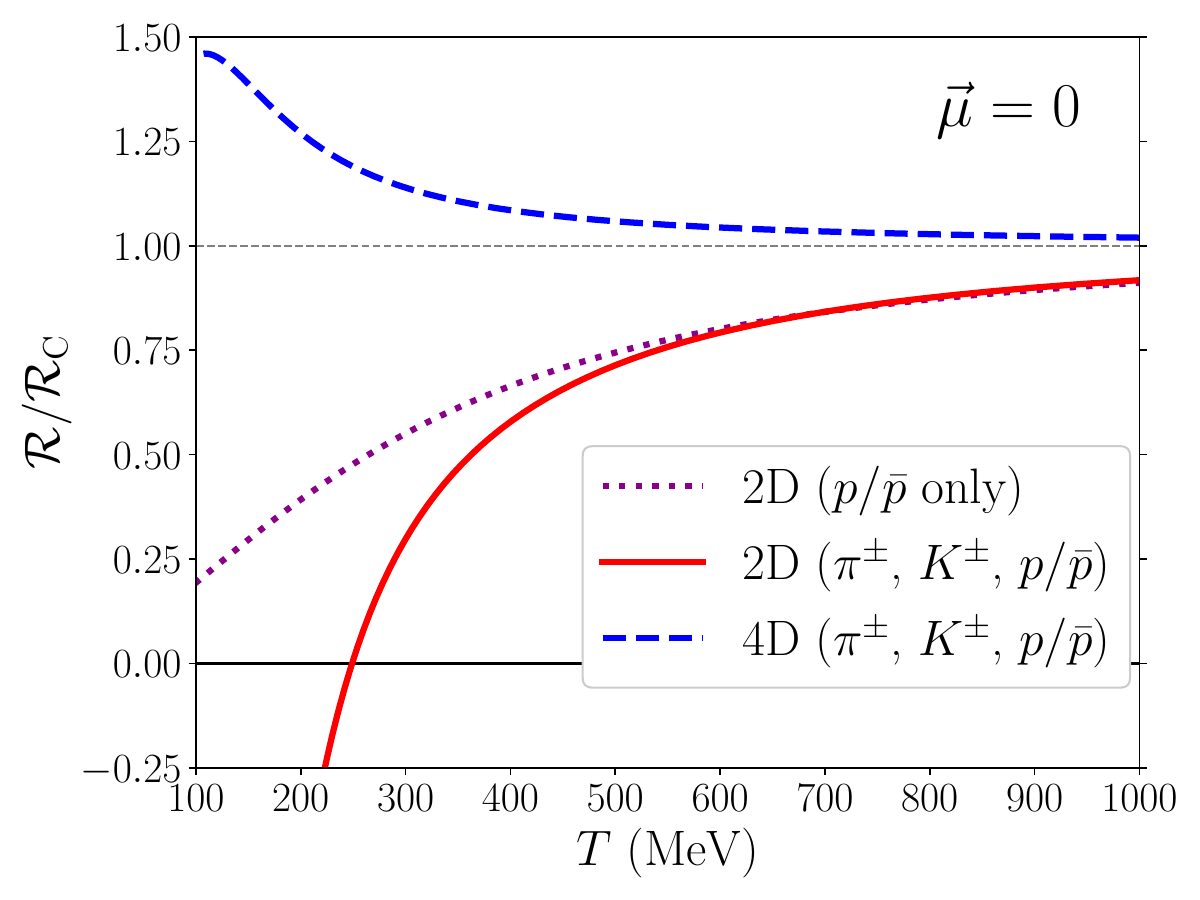}
  \caption{The rescaled scalar curvature $\mR$, normalized by its conformal limit, for three scenarios with differing dimensionalities and hadronic composition: (i), using a 2D analysis ($T$, $\mu_B$) for a system of $p/\bar{p}$ only (dotted, magenta line); (ii), using a similar 2D analysis for the minimal hadron list ($\pi^{\pm}$, $K^{\pm}$, $p/\bar{p}$) in terms of $T$ and $\mu_B$ only (solid, red line); and (iii), using a 4D analysis ($T$, $\mu_B$, $\mu_S$, $\mu_Q$) for the minimal hadron list (dashed, blue line).  A thin, dotted gray line marks the conformal limit.}
  \label{Fig:pi_K_p_analysis}
\end{figure}

\begin{figure}%[h!]
  \centering
  \includegraphics[width=\linewidth]{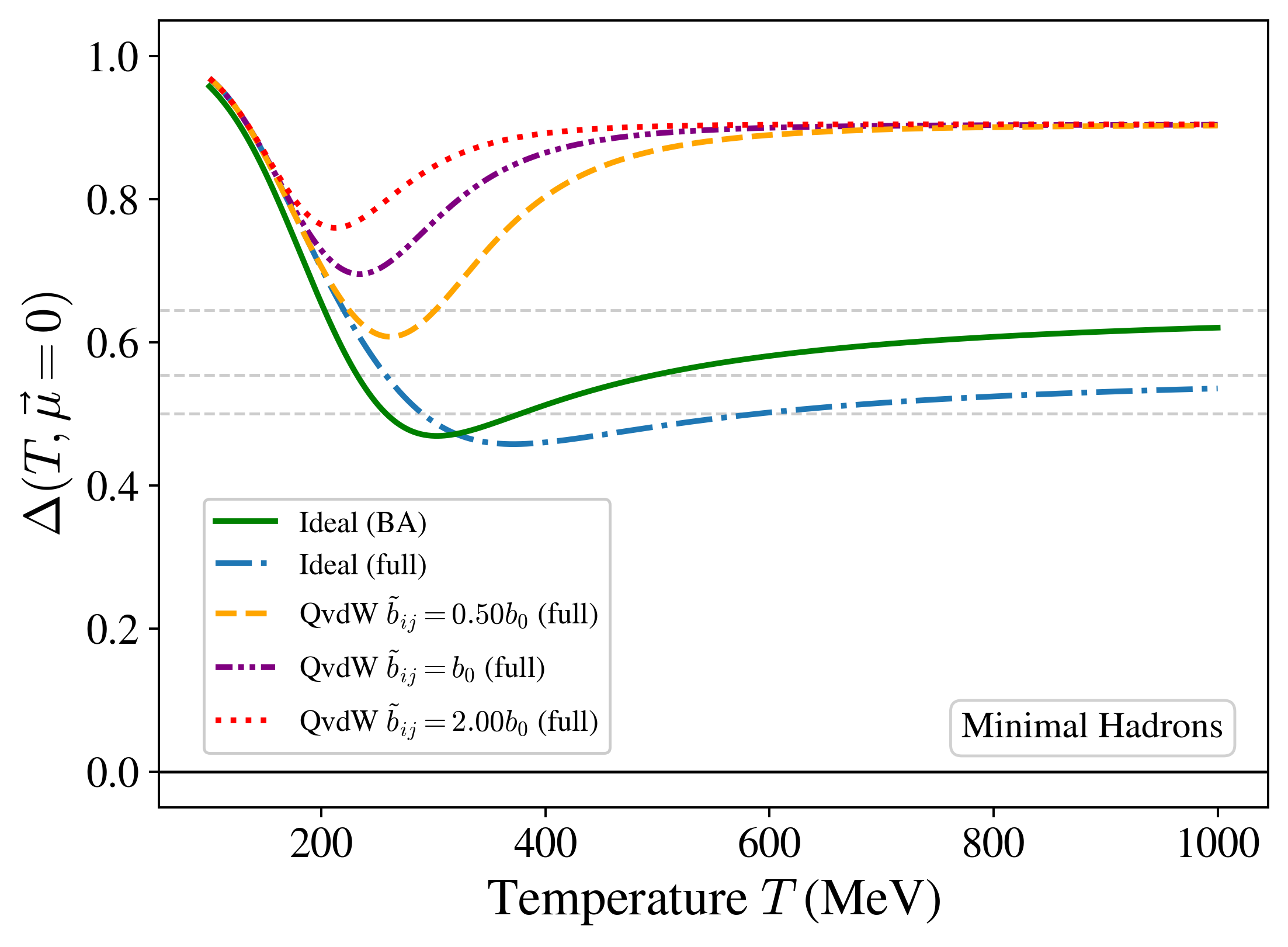}\\
  \caption{The diagonalities for different EoSs using a minimal hadron list (MHL) are compared along the $T$-axis at $\vec\mu = 0$.  Along with the Ideal HRG in the Boltzmann approximation (solid green), we show the Ideal HRG with the full sum over quantum degeneracy effects in Eq.\eqref{p_Ideal_HRG} (blue line with single long dash followed by single dot), and the analogous curves for the Quantum van der Waals (QvdW) model with different values for the excluded volume parameter: $\tilde{b}_{ij} = 0.50 b_0$ (dashed orange), $\tilde{b}_{ij} = b_0$ (purple line with single short dash followed by two dots), and $\tilde{b}_{ij} = 2.00 b_0$ (dotted red).  We also plot thin, gray dashed lines to represent $\Delta = 0.5$, $\Delta = 0.5545$, and $\Delta = 0.6448$, where the latter two values correspond to the high-$T$ limits of the Ideal HRG in the full case and using the Boltzmann approximation.}
  \label{Fig:MHLDiagonalityPlot}
\end{figure}

We next consider the diagonality and the rescaled curvature for the MHL discussed in Sec.~\ref{Sec:MHL_EoS} using the Boltzmann approximation.
Despite the comparative simplicity of \eqref{pi_K_p_EoS}, the resulting expressions for $\mR$ in both the 2D and the 4D cases are quite complicated and so we do not reproduce them here.
Instead, we plot in Fig.~\ref{Fig:pi_K_p_analysis} both cases at $\vec\mu = 0$ as functions of $T$.
For reference, we also show the 2D rescaled curvature for a system consisting of $p/\bar{p}$ only.
All curves are normalized to their conformal limits (i.e., setting $m_\pi = m_K = m_p = 0$), which coincide with the Stefan-Boltzmann limit \eqref{rescaled_R_mu=0_ND_SF}.

For the MHL list, we see that the 2D case exhibits a crossing through zero at $T \approx 250$ MeV, whereas the 4D case is positive for all $T$ shown and does not become negative until $T \lesssim 40$ MeV.
By contrast, the reference 2D case (with $p/\bar{p}$ only) exhibits no crossing of zero for any $T$.
We have also checked (although it is not shown here) that the zero-crossing in the 2D MHL case can be removed either by artificially forcing all particle masses to be equal (i.e., setting $m_p = m_K = m_\pi$) or by introducing a spurious dependence of the mesonic partial pressures on $\mu_B$.
Put differently, only when the baryonic masses differ from the mesonic ones ($m_p \neq m_K$ or $m_p \neq m_\pi$) and the $p_K$ and $p_\pi$ are independent of $\mu_B$ does the 2D MHL exhibit a zero-crossing.
These results hold even if one relaxes the Boltzmann approximation and (incorrectly) requires the mesons to obey fermionic statistics, implying that the sign change of $\mR$ persists even when the whole system behaves fermionically.
This shows that the zero-crossing of $\mR$ is not due to a change from predominantly bosonic to fermionic statistics, \textit{per se}, but is due rather to the influence of the mass differences between physical hadrons on the relative hadronic abundances, together with the fact that mesons necessarily do not carry the conserved baryonic charge whose thermodynamic geometry is being analyzed.
The effect is therefore a reflection of how the competition between the system's baryonic and mesonic degrees of freedom evolves with $T$, not a consequence of a switch between attractive and repulsive interactions.

We conclude from this that an $\mR$-crossing through zero need not imply a qualitative change in the interactions of the system \cite{Ruppeiner_2013, Castorina:2018ayy}, but may more generally result from changes to the system's degrees of freedom.
This suggests that the rescaled curvature may offer valuable insights into the microscopic constituents of a system which are complementary to those obtained from other bulk thermodynamic observables.
We will revisit these points in Sec.~\ref{Sec:Results}.

For $\Delta$, one obtains the results plotted in Fig.~\ref{Fig:MHLDiagonalityPlot}.  We note that the results asymptote to a fixed value at high $T$.  One can understand this by noting that the high-$T$ behavior of $\chi$ at $\vec\mu=0$ is given by
\begin{align}
  \chi &= \frac{2T^2}{\pi^2}
    \begin{pmatrix}
      g_p & 0   & g_p \\
      0   & g_K & g_K \\
      g_p & g_K & g_\pi + g_K + g_p
    \end{pmatrix} 
    = \frac{2T^2}{\pi^2}
    \begin{pmatrix}
      2 & 0 & 2 \\
      0 & 1 & 1 \\
      2 & 1 & 4
    \end{pmatrix} , \label{pi_K_p_chi}
\end{align}
The high-$T$ limit of $\Delta$ is then found to be
\[\lim_{T\to \infty}\Delta(T,\vec\mu=0) = \l( \frac{58-27\sqrt{3}}{65} \r)^{1/4} \approx 0.6448\,,\]
in agreement with the behavior shown in Fig.~\ref{Fig:MHLDiagonalityPlot}.
The $T \to \infty$ limit thus reduces to a non-trivial combination of the different hadron degeneracies in the system and hence displays some sensitivity to its degrees of freedom.

One can repeat this analysis without using the Boltzmann approximation by making use of Eq.~\eqref{p_Ideal_HRG}.  Noting that $p/\bar{p}$ has $s=1/2$ and the mesons have $s=0$, $\chi$ is then
\begin{align}
  \chi &= \frac{T^2}{3}
    \begin{pmatrix}
      1 & 0 & 1 \\
      0 & 1 & 1 \\
      1 & 1 & 3
    \end{pmatrix} , \label{pi_K_p_chi_Full}
\end{align}
corresponding to a diagonality of $\Delta \approx 0.5545$.
%
%\[\lim_{T\to \infty}\Delta(T,\vec\mu=0) = \l( \frac{58-27\sqrt{3}}{65} \r)^{1/4} \approx 0.6448\,,\]
%
This matches the results shown in Fig.~\ref{Fig:MHLDiagonalityPlot}.

\subsubsection*{Toy Model Summary}

Taken together, we find that $\mR$ and $\Delta$ provide complementary insights into the onset of conformality and how a system's degrees of freedom change with evolution in $T$ and $\vec\mu$.
We have shown that on its own $\mR$, when $T \to \infty$, cannot distinguish between the conformal limit of a hadron resonance gas and that of a gas of massless and non-interacting quarks and gluons.
The degrees of freedom in these two systems, however, are fundamentally different and easily distinguished by the high-$T$ limits of the diagonality $\Delta$, which tends to a constant finite value for the former and tends to zero for the latter.

Furthermore, one finds that $\Delta$ alone does not strongly discriminate between the lattice and most of the HRG models for temperatures $T \lesssim 150$ MeV, especially since the lattice parameterization itself is matched smoothly to a HRG model in this regime \cite{Noronha-Hostler:2019ayj}.
Even so, as we demonstrate below, $\mR$ exhibits sensitivity to the EoS model even in this regime.
This suggests that $\mR$ and $\Delta$ together may convey important insights into the properties of the nuclear EoS in the transition region.

%In this sense, $\mR$ exhibits similar limitations to the trace anomaly $I$ \eqref{trace_anomaly}, which also tends to zero in the conformal limit.
%Unlike $I$, however, $\mR$ can be evaluated for a subset of conserved charges or for all present in a system, and Fig.~\ref{Fig:pi_K_p_analysis} shows that the \textit{rate} of $\mR$'s convergence to its high-$T$ limit may depend on which charges are included.

%===========================================================================
%===========================================================================
%===========================================================================
\section{Results}
\label{Sec:Results}

%================================
% Diagonality Results
\begin{figure*}
  \centering
  \includegraphics[width=\textwidth]{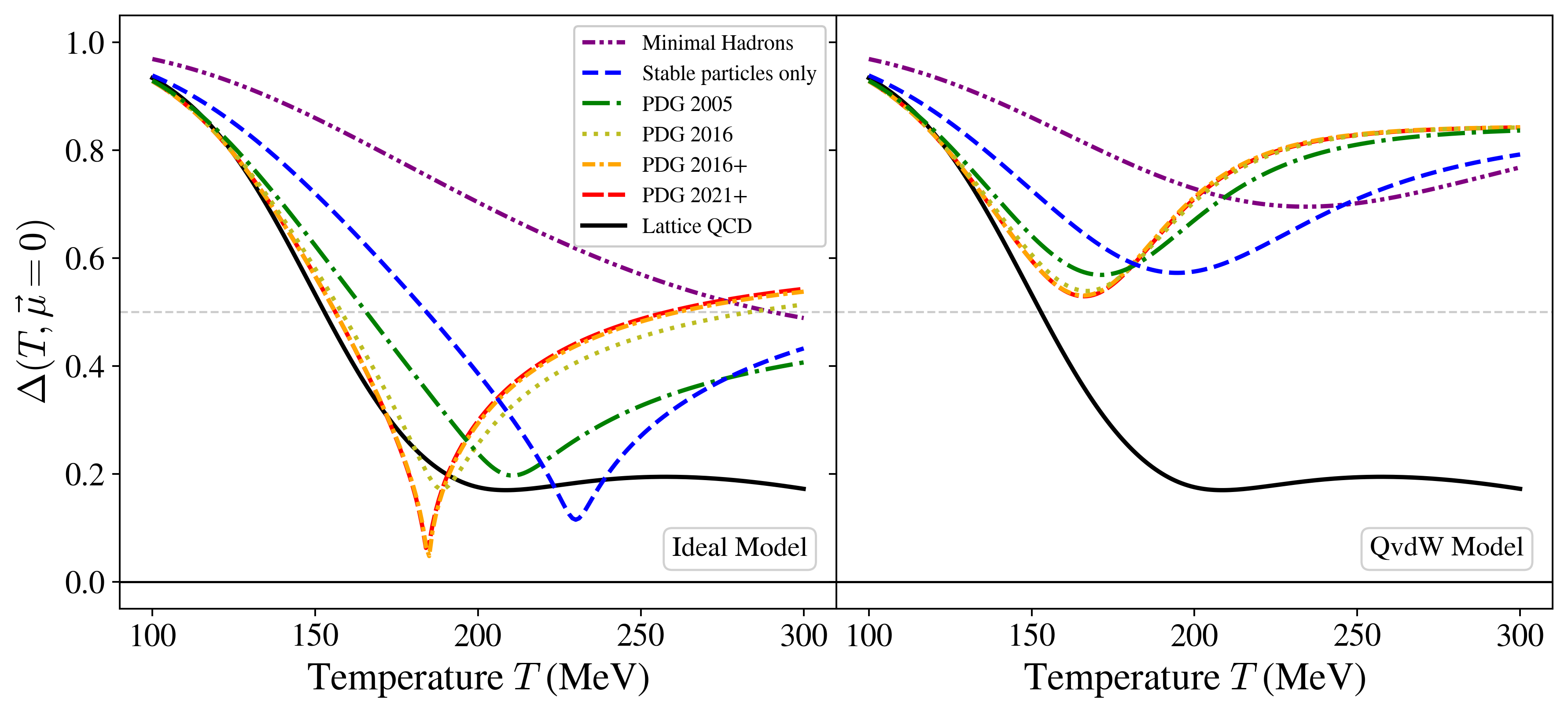}\\
  \caption{The figures above show other diagonality trends containing the same chemical potential constraints in changing temperatures.  The left plot examines the trends of Ideal HRG simulations involving different nuclei lists that contain varying amounts of resonances, while the right plot has the same varying nuclei lists being applied in the context of a QvdW simulation.  The locations of the dips are ordered according to the complexity of the PDG lists, with a greater range of resonances corresponding to a dip at a lower temperature.}
  \label{Fig:DiagonalityPlots}
\end{figure*}

%================================
% Hagedorn Fit Results
\begin{figure}
  \centering
  \includegraphics[width=\linewidth]{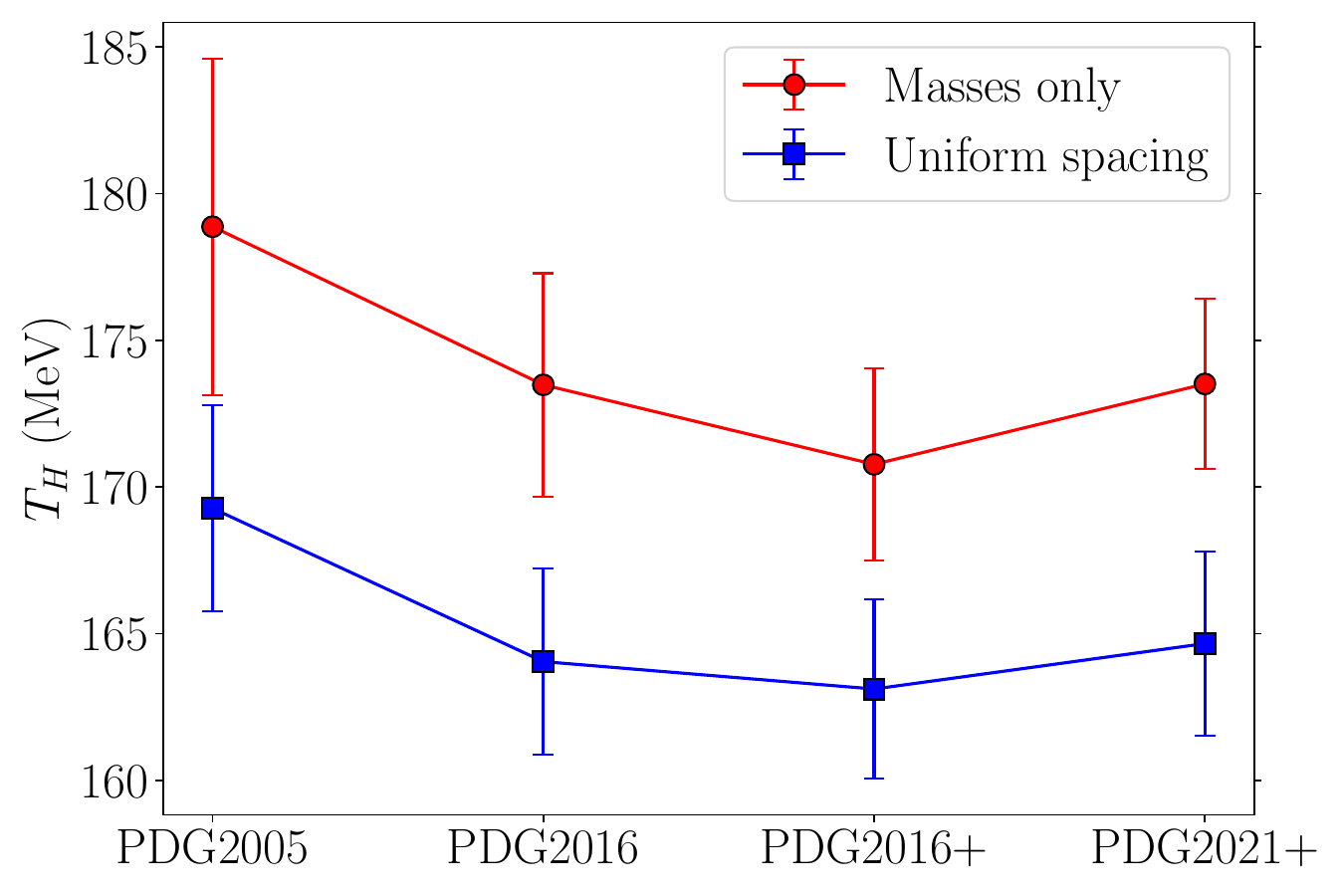}
  \caption{Fits of the PDG lists used in this work to a Hagedorn spectrum of the form Eq.~\eqref{Hagedorn_fit_form}.  Fits using the physical hadron masses (red circles) yield somewhat higher estimates for $T_H$ than those which fit the cumulative distribution function for the spectra, sampled uniformly over the full mass range (blue squares).  Uncertainty estimates are based on the assumption that the fitted points are approximately normally distributed around the fit line.  Further details are given in App.~\ref{App:HagedornFits}.}
  \label{Fig:HagedornFits}
\end{figure}

%================================
% Stable Hadrons List Table

\begin{table}
\begin{tabular}{||l | l||} 
 \hline
 \text{Mesons } & $\pi^0$, $\pi^\pm$, $K^0$, $K^\pm$  \\ 
 \hline
 \text{Baryons } & $n/\bar{n}$, $p/\bar{p}$, $\Lambda^0$, $\Sigma^\pm$, $\xi^0$, $\xi^\pm$, $\Omega^\pm$ \\ [1ex] 
 \hline
\end{tabular}
\caption{The list of stable hadrons used to test how the diagonality and rescaled curvature depend on particle content.}
  \label{Tab:StableParticles}
\end{table}

In the preceding sections we have introduced two observables designed to quantify the degree of independence of different quark flavors and to probe the degrees of freedom in a nuclear system: (i), the diagonality $\Delta$, which measures the proximity of the quark flavor susceptibility matrix $\chi$ to a diagonal form; and (ii), the rescaled thermodynamic curvature $\mR$, which depends only on the number of quark flavors or conserved charges which reach the conformal limit.

In this section we plot several comparisons of the two formalisms discussed above for different hadronic models of nuclear matter.
In order to make robust statements, we systematically vary two separate features of the hadronic models: the Particle Data Group (PDG) list used and, in the case of the QvdW model, the size of the excluded volume parameter $\tilde{b}_{ij}$.

The PDG lists we consider are those from 2005, 2016, 2016+, and 2021+ \cite{Alba:2017mqu, Alba:2020jir, JordiSalinasSanMartinGitHubRepo, SanMartin:2023zhv}.
We also include two abbreviated lists of hadrons, one consisting only of stable hadrons (those listed in Table \ref{Tab:StableParticles}) and the MHL list introduced in Sec.~\ref{Sec:MHL_EoS}.
While these abbreviated lists are not physically realistic descriptions of nuclear matter, they enable clearer insights into the way the diagonality and thermodynamic curvature behave as functions of temperature and how this behavior reflects the underlying state of the system.
Likewise, varying the excluded volume parameter (by either multiplying or dividing the default value of $b_0 = 3.42$ fm$^3$ by a factor of 2) allows us to see how $\Delta$ and $\mR$ can be sensitive to hadronic model parameters, and we further compare these models with the results for lattice QCD.
We begin first with the diagonality and then discuss the thermodynamic curvature.

%===========================================================================
\subsection{Diagonality}
\label{Sec:Results:Diagonality}

We begin by showing the effects of varying the PDG list used in the HRG models.
Fig.~\ref{Fig:DiagonalityPlots} shows the diagonality of the Ideal and QvdW hadronic models, together with lattice, for the different PDG lists mentioned above.
We notice several key features in these plots.

\textit{Crossover in Lattice QCD}: First, we observe that the lattice QCD diagonality begins close to 1 at temperatures below the chiral crossover and decreases steadily until $T \approx 200$ MeV.
This shows that the susceptibility matrix in the flavor basis is far from being diagonal as $T \to 0$, but becomes more diagonal as the temperature is increased.
This implies that quark dynamics become increasingly well approximated by the Stefan-Boltzmann limit as well, which is reasonable since the lattice QCD EoS \cite{Noronha-Hostler:2019ayj} was constructed to do precisely this.
We observe that $\Delta = 0.5$ at approximately $T_{\Delta = 0.5} = 153$ MeV, in reasonably good agreement with the chiral crossover temperature of 156.5 (1.5) MeV \cite{HotQCD:2018pds}.
In a rough sense, $T_{\Delta = 0.5}$ is the temperature above which quarks behave `more independently than not.'
Nevertheless, the diagonality clearly does not drop immediately to zero above $T_{\Delta = 0.5}$, as would be expected from the formation of an asymptotically free plasma, but remains of order 0.2 at least up to $T = 300$ MeV.
The onset of deconfinement in lattice QCD is a gradual process which continues at least up to $2T_\mathrm{pc}$.

Second, comparing the lattice results to those of the various hadronic models in Fig.~\ref{Fig:DiagonalityPlots} reveals that more complete PDG lists produce a closer approximation to the lattice QCD diagonality.
This holds for both the Ideal and the QvdW models, for which one sees that those lists constructed in 2016 or later remain the closest to the lattice up until the crossover temperature.
The same holds true, albeit to a lesser degree, if one compares the 2016 list to the 2016+ and 2021+ lists (where the latter pair of lists contains a larger number of missing strange resonances): while the QvdW models are only weakly affected in this case, the Ideal models show a detectable sensitivity to the list content which slightly favors the more extended lists.
This in turn shows that utilizing the most extensive list of hadronic resonances available is essential for reproducing observables like the diagonality \cite{Lo:2015cca, Alba:2017mqu, SanMartin:2023zhv}.

Third, all hadronic models exhibit a dip in their diagonalities in the crossover region, followed by a partial rise at larger $T$ and an eventual saturation at a constant value somewhat less than unity.
This qualitative behavior is explained by the competition between different contributions to the total susceptibilities from hadronic species with different masses, as was noted above in Sec.~\ref{Sec:MHL_EoS}.
At low $T$, the charged hadrons with the smallest masses (e.g., pions) tend to dominate the corresponding susceptibilities.
As $T$ increases, however, fluctuations from heavier resonances become comparatively more important and drive $\chi$ toward a more diagonal form, causing $\Delta$ to begin to decrease.
For the Ideal minimal hadron system (Fig.~\ref{Fig:MHLDiagonalityPlot}), $\Delta$ continues to decrease until the proton susceptibility is approximately half as large as the pion susceptibility, which leads to $\chi_{ud} \approx 0$ and hence a minimum value for $\Delta$.
At larger $T$, $\Delta$ continues to increase from its minimum until the pion and proton susceptibilities become comparable in magnitude, causing $\Delta$ to saturate at a constant value which is a function only of the conserved charges carried by each hadron (cf.~Eqs.~\eqref{pi_K_p_chi} and \eqref{pi_K_p_chi_Full}).
The diagonality therefore reflects in a non-trivial way the hadronic content of a nuclear EoS and how this content changes with $T$.

Fourth, one observes a hierarchy in the temperatures at which the diagonality crosses the $\Delta = 0.5$ line and/or at which it reaches its minimum value.
This effect is most pronounced in the Ideal models, but the QvdW models exhibit a similar trend.
Specifically, we find that those models with the most complete PDG lists reach their minima first (i.e., at smaller $T$).
        
To better understand this `temperature hierarchy,' we use a similar fitting procedure to that in \cite{SanMartin:2023zhv} to extract an estimate of $T_H$ for the each PDG list used in our analysis.
The fit results are shown in Fig.~\ref{Fig:HagedornFits}, with the red circle and blue squares denoting different ways of extracting $T_H$, as discussed in App.~\ref{App:HagedornFits}.
One finds that the PDG 2005 list yields a larger $T_H$ than the other three PDG lists, which all yield consistent estimates for $T_H$ within the estimated uncertainties.
This hierarchy qualitatively matches the ordering of minima in $\Delta$ visible in Fig.~\ref{Fig:DiagonalityPlots}, suggesting that $\Delta$ is sensitive to the particle content and, more generally, the degrees of freedom in the system.

In Fig.~\ref{Fig:MHLDiagonalityPlot}, one can observe how $\Delta$ also depends on the excluded volume parameter $\tilde{b}_{ij} = 0.50 b_0$, $1.00 b_0$, and $2.00 b_0$ (with the default value $b_0 = 3.42$ fm$^3$).
First, noting that the Ideal HRG corresponds to taking $b \to 0$, we see that decreasing $b$ leads to diagonalities which converge to the Ideal case.
For $b>0$, the dip in the diagonality is less pronounced than in the Ideal case.
This is easily understood from the fact that a finite $b>0$ suppresses baryon susceptibilities relative to the Ideal HRG \cite{Vovchenko:2017zpj}.
Hence, in the QvdW HRG using the MHL (Fig.~\ref{Fig:DiagonalityPlots}), increasing $b$ produces smaller proton susceptibilities relative to those for pions, causing the minimum in $\Delta$ to increase.

The preceding observations demonstrate that the diagonality quantifies the degree of independence of quarks and also constitutes a sensitive probe of the EoS.
We demonstrate next that the thermodynamic curvature serves a similar purpose and conveys important complementary insights into the EoS of nuclear matter.

%===========================================================================
\subsection{The Rescaled Curvature}
\label{Sec:Results:RescaledCurvature}

%================================
% Thermodynamic Geometry Results
\begin{figure*}%[!htpb]
  \centering
  \includegraphics[width=\textwidth]{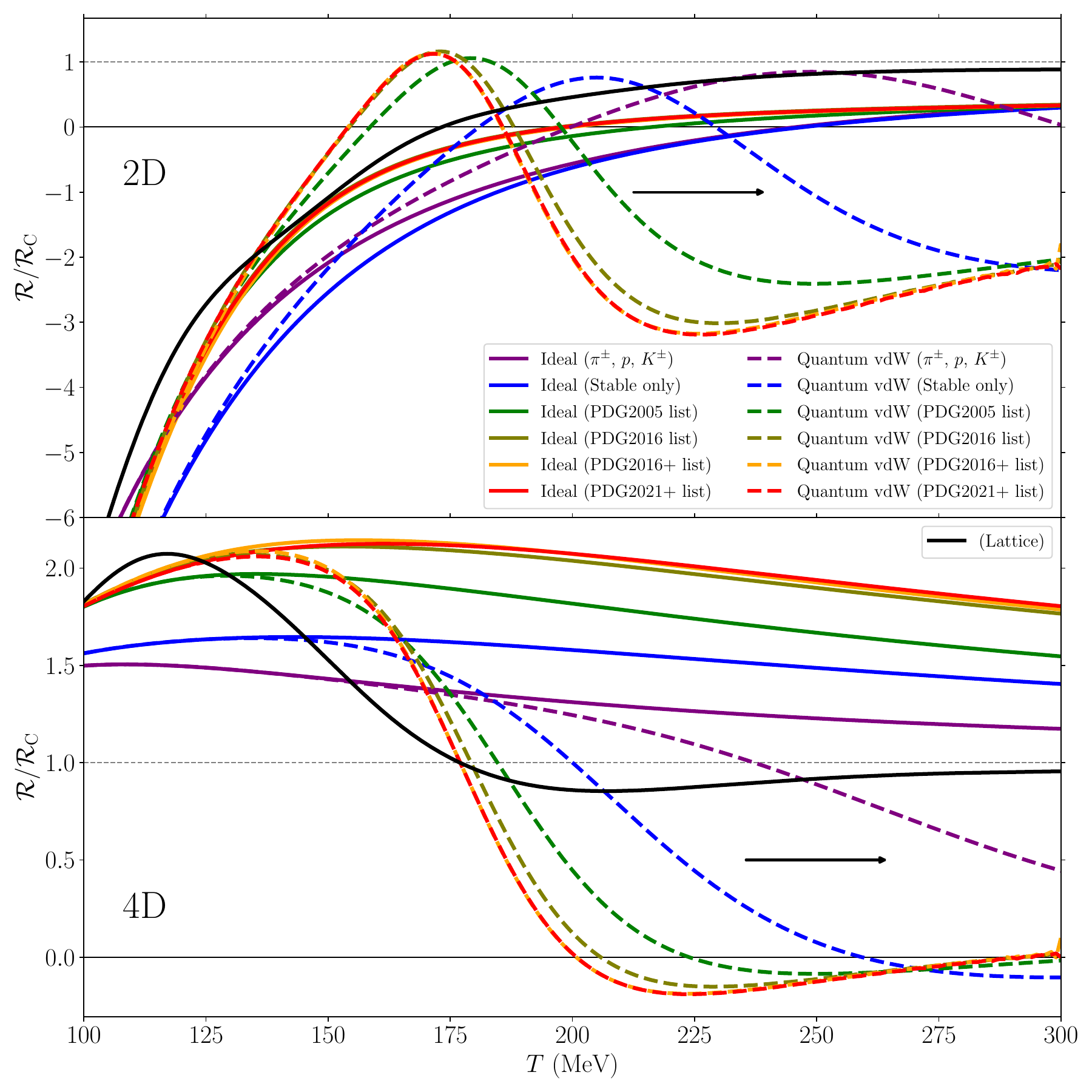}
  \caption{The Ricci scalar multiplied by the entropy density $\mR(T,\vec\mu = 0)$, normalized to the Stefan-Boltzmann limit, for Lattice QCD (solid, black lines) compared with several HRG models.  Both Ideal (solid, colored lines), and Quantum van der Waals (dashed, colored lines) are shown in the 2D (top) and 4D (bottom) cases.  Different colors represent the different hadron lists used, including PDG2005 (green), PDG2016 (olive), PDG2016+ (orange), PDG2021+ (red), a list consisting only of stable particles (blue), and the `minimal hadron set' list consisting only of $\pi^\pm$, $K^\pm$, and $p/\bar{p}$ (purple).  The solid black arrows emphasize the ``temperature hierarchy" which parallels those in Figs.~\ref{Fig:MHLDiagonalityPlot} and \ref{Fig:DiagonalityPlots} and qualitatively agrees with the hierarchy visible in Fig.~\ref{Fig:HagedornFits}.  The lattice results clearly approach their conformal limit values by roughly $T \approx 300$ MeV, while the HRG models do not converge until much later.}
  \label{Fig:R_times_s_vs_T_compare_PDG}
\end{figure*}

In Fig.~\ref{Fig:R_times_s_vs_T_compare_PDG}, we plot the rescaled curvature $\mR$ at $\vec\mu=0$ as a function of $T$ for several different EoS models and particle lists used above.
We show results for the Ideal (solid) and QvdW (dashed) HRG models, computed for the PDG2005 (green), PDG2016 (olive), PDG2016+ (orange), PDG2021+ (red), a list consisting only of stable particles (blue), and the MHL introduced in Sec.~\ref{Sec:MHL_EoS} (purple).
We compare both the 2D and 4D cases, and compare also with the corresponding lattice curvature as solid black curves \cite{Noronha-Hostler:2019ayj}.  As with the diagonality above, we make several observations.

First, the rescaled curvature behaves in qualitatively different ways in the 2D and 4D cases.
Similarly to the case of the MHL particle list above, in the 2D case we find all EoSs considered cross zero at least once in the vicinity of $100 \lesssim T \lesssim 200$ MeV.
It has been proposed to interpret this as the occurrence of a crossover from predominantly attractive to predominantly repulsive interactions as might occur with the onset of deconfinement \cite{Castorina:2018ayy, Castorina:2018gsx, Castorina:2019jzw, Castorina:2020vbh}.
However, as we argued above in Sec.~\ref{Sec:MHL_EoS}, the zero-crossing of $\mR$ is more plausibly attributed to the system's evolving particle content, rather than to any qualitative change in its interactions.
Like the diagonality, the rescaled curvature thus reflects the system's underlying degrees of freedom and how these change with $T$.

Second, we observe a similar ordering in the temperature dependences (indicated schematically with black arrows) of different PDG lists as was observed in the diagonalities of Fig.~\ref{Fig:DiagonalityPlots}.
In the 2D QvdW cases, for instance, we note that the locations of the peaks are ordered similarly to the corresponding Hagedorn temperatures extracted from each list, as are the $\mR = 0$ crossings in the range 175 MeV $\lesssim T \lesssim 225$ MeV in the Ideal HRG models.
Similarly, in the 4D cases, the places where the QvdW models cross the Stefan-Boltzmann limit and zero again reflect the same ordering.
In general, we find that the PDG2016+ and PDG2021+ lists reach these features first, followed by the PDG2016 list, and followed finally by PDG2005, the stable particle list, and the minimal particle list.
As with the diagonalities shown above, this underscores a relation between an effective Hagedorn-like temeperature characterizing the transition region for a given EoS and the bulk thermodynamics which the rescaled curvature encapsulates.

Third, we see clearly that the Ideal HRG curves eventually converge to the conformal limit, although even the most complete PDG lists do so at a much slower rate than do the lattice curves.
This is consistent with the observations made above that both the Stefan-Boltzmann EoS \eqref{SB_limit_massless} and the high-$T$ limit of the Ideal HRG \eqref{pi_K_p_EoS_massless} eventually tend to a conformal limit \eqref{EoS_conformal_limit_general}.

The lattice curvatures only converge to within 5\% of the conformal limit for $T \gtrsim 275$ MeV.
This suggests that quarks continue to exhibit non-negligible cross-flavor correlations well above the chiral crossover temperature, which is consistent with observations based on other observables like the Polyakov loop \cite{McLerran:2026dio}.  As with $\Delta$, $\mR$ suggests a gradual onset to deconfinement above $T_\mathrm{pc}$.

%===========================================================================
%===========================================================================
%===========================================================================
\section{Conclusions}
\label{Sec:Conclusions}

%In this work, we have demonstrated how novel insights into the equation of state (EoS) of nuclear matter can be generated by constructing quantities such as the rescaled thermodynamic scalar curvature and the diagonality of the conserved charge susceptibility matrix, and by using these observables as quantitative probes of the EoS's phase structure.  We find that both quantities, when applied to the full $BSQ$ susceptibility matrix, provide complementary perspectives on the state and properties of nuclear matter as functions of temperature $T$ and conserved charge chemical potentials $\mu_{B,S,Q}$.  Our main conclusions are as follows:

In this work, we introduced two new observables -- the diagonality $\Delta$ and the rescaled curvature $\mR$ -- designed to function as sensitive, quantitative probes of the nuclear EoS and its phase diagram.
Both quantities, when applied to the full susceptibility matrix $\chi$, provide complementary perspectives on the state and properties of nuclear matter as functions of temperature $T$ and conserved charge or quark flavor chemical potentials $\vec\mu$.

Our main conclusions are as follows:

\begin{itemize}
  \item The chiral transition temperature coincides very nearly with the point at which the lattice diagonality $\Delta = 0.5$.  One may roughly interpret this as the point at which quarks begin to behave more like independent particles than not.  However, the onset of full deconfinement is a gradual process which remains incomplete at least until $T \approx 2T_\mathrm{pc}$.
  \item The characteristic dip seen in the diagonality for HRG models can be understood in terms of the competing temperature dependences of the susceptibilities.
  At small $T$, all hadron densities are small and comparable, leading to a diagonality of 1.
  As $T$ increases, the lightest particles (pions) increase more rapidly than heavier particles, leading to strong fluctuations and a more diagonal susceptibility matrix.
  Eventually, however, as $T$ continues to increase, all particles become relativistic and contribute comparably to the susceptibilities, leading to a rise in the diagonality at high $T$.
  
  Furthermore, \textit{decreasing} the excluded volume in the QvdW model renders the chiral transition dip \textit{more} pronounced.
  This is due to a two-fold effect:
    (i), a finite excluded volume ($\tilde{b}_{ij}>0$) reduces the rate at which baryon susceptibilities can grow, making the dip less pronounced; and
    (ii), decreasing $\tilde{b}_{ij}$ toward zero more closely approximates the Ideal HRG which has $\tilde{b}_{ij}=0$ identically.
  $\Delta$ is therefore also sensitive to HRG model parameters.
  \item The rescaled curvature crosses zero in the 2D case, but not necessarily in the 4D case.
  This reflects the changing particle content and thermodynamics of conserved quantities present in the system and is not primarily related to changing statistics or interactions.
  $\mR$ is thus a sensitive probe of an EoS's degrees of freedom and reflects how they evolve with $T$ and $\vec\mu$.
  \item Comparing the diagonality and $\mR$ for different PDG lists illustrates how the inclusion of additional resonances is essential for accurately modeling the HRG phase of QCD matter.
  One observes an ordering in the temperature dependencies of all quantities which parallels the ordering of the Hagedorn temperatures in the HRG models considered.
  The two quantities we have discussed here (i.e., $\Delta$ and $\mR$) therefore provide connections between the Hagedorn-like scales characterizing resonance mass spectra and bulk thermodynamic quantities, with the added advantage that they can be evaluated for any EoS, even those without well defined particle lists.
\end{itemize}

The research presented here opens up several avenues for further exploration of the QCD EoS.

First, both the diagonality and the rescaled curvature should be studied at finite conserved charge densities and chemical potentials, particularly to assess their sensitivity to critical point effects.
Doing so could also provide an alternative definition of the QCD transition curve as a function of chemical potential and might help to elucidate the nature of the quark-hadron transition at finite density.
Similarly, it will also be important to study other specific models of nuclear matter using our observables.
This would be especially valuable for identifying quantitative signatures of the `stringy fluid' phase which has been conjectured to exist above $T_\mathrm{pc}$.

Our findings suggest that the observables introduced here could potentially be used to discriminate between competing models of nuclear matter in a given regime.
Specifically, the diagonality and rescaled curvature, given their sensitivity to the degree of independence of quark flavors and the underlying degrees of freedom, could potentially be used to search for evidence of the recently conjectured `stringy fluid' discussed above \cite{Ratti:2011au, McLerran:2026dio}.
We plan to pursue these questions in future studies.

One could also consider developing the observables themselves further.
For example, the diagonality is only a single statistic which measures how diagonal the quark flavor susceptibility matrix $\chi$ is.
In particular, it does not explicitly convey the behavior of $\chi$'s eigenvalues or eigenvectors themselves, which could contain additional valuable information about the thermodynamic state of a given nuclear system.
Another possibility would be to consider additional quark flavors in the diagonality, e.g., charm quark susceptibilities.

Similarly, it would be interesting to generalize the definition of the rescaled curvature to account finite charge densities, which might enable clearer insights into the properties of QCD matter in other regimes currently inaccessible to lattice calculations.

One further application of our results here would be to study the evolution of the diagonality and rescaled curvature in strongly non-equilibrium systems such as heavy-ion collisions which simultaneously evolve multiple conserved charge densities \cite{Plumberg:2024leb, Pala:2025qoa}.
Given the insights they provide into the nature of the QCD transition region, these observables could yield fruitful insights which would help to improve the phenomenological treatment of freeze out and particlization in heavy-ion collision simulations.

\begin{acknowledgments}
  We thank Joshua Bowman for helpful comments about different ways to quantify the diagonality of a matrix.  We gratefully acknowledge support for this project by the Academic Year Undergraduate Research Initiative (AYURI) and Undergraduate Research Fellowships (URFs) at Pepperdine University.  C.P. thanks Jacquelyn Noronha-Hostler and Joseph Kapusta for helpful comments and suggestions, and Jordi Salinas San Mart\'in for valuable guidance on extracting the Hagedorn temperature from a list of hadrons.
\end{acknowledgments}

\appendix

\section{Hagedorn Fitting Procedure}
\label{App:HagedornFits}

To assign a Hagedorn temperature $T_H$ to a given set of hadrons such as one of the PDG lists employed here, we first define the cumulative distribution function of the hadron set and the corresponding functional to which this distribution is fit in order to extract $T_H$.
Since the mass spectrum of the hadron set \eqref{PDG_mass_spectra} does not exactly follow the exponential distribution \eqref{Hagedorn_fit_form}, there is necessarily some ambiguity in how the fit is carried out.

First, we fit the cumulative distribution function of the masses in a given list, defined by
\begin{align}
  N_\text{exp}(m) = \sum_{i \in \text{hadrons}} \Theta\l( m - m_i \r) \label{PDG_mass_spectra}
\end{align}
to a function form given by \cite{SanMartin:2023zhv}
\begin{align}
  N_\text{fit}(m) = \frac{A \exp(m/T_H)}{(m^2 + m_r^2)^{5/4}} \label{Hagedorn_fit_form}
\end{align}
where $A$ is a normalization factor, $T_H$ is the Hagedorn temperature, $m_r = 500$ MeV is a fixed mass scale, and $\Theta(x)$ is the Heaviside step function.

The red circles represent fits to the hadron mass spectrum sampled only at the physical PDG particle masses in the range 0.5 - 1.68 GeV.
By contrast, the blue squares represent a uniform sampling of points in the same mass range, but sampling the spectrum with a uniform mass spacing of $dm = 0.01$ GeV.
In both cases, the irregularities of the physical mass spectrum are used to estimate the uncertainties on the extracted $T_H$.

To assess the uncertainty in the extracted value for $T_H$, we consider two different prescriptions for fitting the spectra.
The first prescription samples the cumulative distribution function for the masses of the hadron set only at the physical masses themselves.
The second prescription first determines the fit range (in this case, from masses of 0.5 to 1.68 GeV) and then samples the cumulative distribution function uniformly over this interval.
%The sample point sets corresponding to the first (red points) and second (blue points) prescriptions are shown for the PDG2021+ list in Fig.~\ref{}, along with the corresponding fit lines (solid lines).
In both cases, the scatter of points about the corresponding fit line provides an indication of the quality of the fit.
Furthermore, comparing the two prescriptions side-by-side provides a complementary estimate of how sensitively the extracted $T_H$ depends on the details of the fit procedure.

%===============================================================
\section{Rescaled curvature for general $N_f$ in massless, non-interacting limit}
\label{App:RicciCurvatureGeneral}

In \cite{Ruppeiner:1995zz} it was pointed out that multiple equivalent coordinate systems are available for computing thermodynamic geometry, and that some coordinates may be better suited to certain problems than others.  Here, we review the use of coordinates which are more natural for the GCE and show how the coordinates can be applied to the thermodynamic geometry.

In the GCE, the independent variables are the temperature $T$ and the assorted chemical potentials $\vec\mu = (\mu_1, \ldots, \mu_N)$.  Following \cite{Ruppeiner:1995zz}, one may define the following metric:
%
%\begin{widetext}
\begin{align}
  g \equiv - \frac{1}{T}
    \begin{pmatrix}
      %---------------------------------------------------------------
      \chi_{TT}     & \chi_{T\mu_1}     & \cdots & \chi_{T\mu_N}     \\
      %---------------------------------------------------------------
      \chi_{T\mu_1} & \chi_{\mu_1\mu_1} & \cdots & \chi_{\mu_1\mu_N} \\
      %---------------------------------------------------------------
      \vdots        & \vdots            & \ddots & \vdots            \\
      %---------------------------------------------------------------
      \chi_{T\mu_N} & \chi_{\mu_1\mu_N} & \cdots & \chi_{\mu_N\mu_N}
      %---------------------------------------------------------------
    \end{pmatrix} \label{metric4D_definition_ALT}
\end{align}
%\end{widetext}
%
Here we have generalized the notation used in Eq.~\eqref{charge_susceptibilities} to include derivatives with respect to both $T$ and $\vec\mu$.  Eq.~\eqref{metric4D_definition_ALT} is immediately recognized as (proportional to) the full susceptibility matrix in either the flavor or charge basis.
We may therefore think of $T$ as a `timelike' coordinate and the $\vec\mu$ as $N$ `spacelike' coordinates.
Since $\chi$ is readily available when working in the GCE, this formalism is generally preferable for analyses of the QCD transition region.

%===============================================================

As an application of this alternate coordinate system, we now show how the Ricci Scalar $R$ depends on the number of massless, non-interacting quark flavors (or conserved charges) present in the equation of state and which are included in the geometric analysis.
We consider only the case with $\vec\mu=0$, corresponding to the `spatial' origin of the thermodynamic phase space.
First, we note that, since the thermodynamic coordinates are $(T,\vec\mu)$, the chemical potentials represent the `spatial' dimensions of the thermodynamic phase space and we are therefore interested in the behavior of $R$ at the `spatial' origin of this space.
Moreover, all such spatial dimensions are completely equivalent to each other since all quarks behave identically in the Stefan-Boltzmann limit.

To simplify our notation, we define $d \equiv N_f$ and rewrite the EoS with arbitrary coefficients:
\begin{align}
  p = T^4 \l[ \kappa
          + \sum_{f=1}^d
             \l( a
                + b \l( \frac{\mu_f}{T} \r)^2
                + c \l( \frac{\mu_f}{T} \r)^4
             \r)
           \r] \label{EoS_general_SB}
\end{align}

The metric \eqref{metric4D_definition_ALT} in this case has the form
\begin{widetext}
\begin{align}
  g \equiv -
    \begin{pmatrix}
      %---------------------------------------------------------------
      12 T(\kappa + d a) + \frac{2b}{T} \sum_{f=1}^d \mu_f^2 & 4 b \mu_1 & \cdots &  4 b \mu_d \\
      %---------------------------------------------------------------
      4 b \mu_1 & 2 b T + \frac{12 c}{T} \mu_1^2 & \cdots &  0 \\
      %---------------------------------------------------------------
      \vdots & \vdots & \ddots &  \vdots \\
      %---------------------------------------------------------------
      4 b \mu_d & 0 & \cdots &  2 b T + \frac{12 c}{T} \mu_d^2
      %---------------------------------------------------------------
    \end{pmatrix} \label{metric4D_definition_ALT2}
\end{align}
\end{widetext}
Next, we use the fact that $g_{\mu\nu}$ is diagonal at $\vec\mu = 0$ to write $R$ as
\begin{align}
  R
    &= g^{\mu\nu} R_{\mu\nu}
    = g^{00} R_{00} + \sum_{i=1}^d g^{ii} R_{ii} \\
    &= g^{00} R_{00} + d \frac{R_{ii}}{g_{ii}} \label{EoS_Ricci_scalar}
\end{align}
where in the final step we have made use of the equivalence of the various `spatial' dimensions.
We therefore need only evaluate the diagonal elements of the Ricci tensor, which are given by
\begin{align}
  R_{\alpha\alpha}
    &= \partial_\lambda \Gamma^\lambda_{\alpha\alpha}
     - \partial_\alpha \Gamma^\lambda_{\alpha\lambda}
     + \Gamma^\lambda_{\lambda\sigma} \Gamma^\sigma_{\alpha\alpha}
     - \Gamma^\lambda_{\alpha\sigma} \Gamma^\sigma_{\lambda\alpha}
\end{align}
where no summation is implied over the $\alpha$ index.
Using standard computer algebra software to evaluate derivatives, we find when $\vec\mu = 0$ that the only non-vanishing Christoffel symbols are
\begin{align}
  \Gamma^0_{00} = \Gamma^i_{i0} = \Gamma^i_{0i} = \frac{1}{2T}, \quad \Gamma^0_{ii} = \frac{b}{4T(a d+\kappa)}
\end{align}
and likewise the only non-vanishing derivatives of Christoffel symbols are (again with no indices summed over)
\begin{align}
  \partial_0 \Gamma^0_{00} = \partial_0 \Gamma^i_{i0} = \partial_0 \Gamma^i_{0i} &= -\frac{1}{2T^2}, \\
  \partial_0 \Gamma^0_{ii} &= -\frac{b}{4T^2 (a d+\kappa)} \\
  \partial_i \Gamma^i_{00} = -\frac{2}{T^2},\, \partial_i \Gamma^i_{ii} &= \frac{1}{T^2} \l( \frac{6c}{b} - \frac{b}{2(a d + \kappa)} \r) \\
  \partial_i \Gamma^i_{jj} &= -\frac{b}{2T^2 (a d+\kappa)} \, (i \neq j)
\end{align}
This yields the Ricci tensor's diagonal components:
\begin{align}
  R_{00} = -\frac{3d}{2T^2},\quad R_{ii} = \frac{b(1-3d)}{8 T^2 (a d + \kappa)}
\end{align}
Substituting back into Eq.~\eqref{EoS_Ricci_scalar} then finally gives
\begin{align}
  R
    &= \l( -\frac{1}{12 T(a d +\kappa)} \r) \l( -\frac{3d}{2T^2} \r) \nonumber\\
    &+ d\, \l( -\frac{1}{2 b T} \r) \l( \frac{b(1-3d)}{8 T^2 (a d + \kappa)} \r) \\
    &= \frac{d(3d+1)}{16(ad+\kappa)T^3}
\end{align}
Finally, multiplying this result by the entropy density
\begin{align}
  s(T, \vec\mu) &= \frac{\partial p}{\partial T} = 4(a d+\kappa) T^3
\end{align}
and making the replacement $d \to N_f$ yields Eq.~\eqref{rescaled_R_mu=0_ND_SF}.
We have checked by explicit computation up to $N_f = 6$ that \eqref{rescaled_R_mu=0_ND_SF} yields the correct value of $\mR$ at $\vec\mu = 0$.

% Place the bibliography commands at the end of the document
%\bibliographystyle{unsrt}        % Determines how the reference list looks (e.g., plain, alpha, apalike)
\bibliography{references,not_inspirehep}

\end{document}